\magnification=\magstephalf
\hsize=13cm
\vsize=20cm
\overfullrule 0pt
\baselineskip=13pt plus1pt minus1pt
\lineskip=3.5pt plus1pt minus1pt
\lineskiplimit=3.5pt
\parskip=4pt plus1pt minus4pt

\def\negenspace{\kern-1.1em}


\newcount\secno
\secno=0
\newcount\susecno
\newcount\fmno\def\z{\global\advance\fmno by 1 \the\secno.
                       \the\susecno.\the\fmno}
\def\section#1{\global\advance\secno by 1
                \susecno=0 \fmno=0
                \centerline{\bf \the\secno. #1}\par}
\def\subsection#1{\medbreak\global\advance\susecno by 1
                  \fmno=0
       \noindent{\the\secno.\the\susecno. {\it #1}}\noindent}

\def\sqr#1#2{{\vcenter{\hrule height.#2pt\hbox{\vrule width.#2pt
height#1pt \kern#1pt \vrule width.#2pt}\hrule height.#2pt}}}

\newcount\refno
\refno=1
\def\y{\the\refno}
\def\myfoot#1{\footnote{$^{(\y)}$}{#1}
                 \advance\refno by 1}

\def\newref{\vskip 0.1pc 
            \hangindent=2pc
            \hangafter=1
            \noindent}
\def\neq{\hbox{$\,$=\kern-6.5pt /$\,$}}



\newcount\secno
\secno=0
\newcount\fmno\def\z{\global\advance\fmno by 1 \the\secno.
                       \the\fmno}
\def\sectio#1{\medbreak\global\advance\secno by 1
                  \fmno=0
       \noindent{\the\secno. {\it #1}}\noindent}

\centerline{\bf ANALOGUE OF BLACK STRING IN THE YANG--MILLS GAUGE THEORY}
\bigskip\bigskip

\centerline{Yuri N.~Obukhov$^{*}$}
\bigskip  
\bigskip
\centerline{Institute for Theoretical Physics, University of Cologne,}
\centerline{D--50923 K\"oln, Germany}


\bigskip  
\bigskip
\centerline{\bf Abstract}
\medskip
The classical Yang--Mills equations are analyzed within the geometrical
framework of an effective gravity theory. Exact analytical solutions are
derived for the cylindrically symmetric configurations of the coupled
gauge and isoscalar fields. It turns out that there is an infinite family
of solutions parametrized by two real parameters, one of which determines
the asymptotic behavior of fields near the symmetry axis and in infinity, 
while the second locates the singularity. These configurations have a simple
pole at a finite value of the radial coordinate, and physically they
represent ``thick string''-like objects which possess the confinement
properties. It is demonstrated that the particles with gauge charge cannot 
move classically and quantum mechanically out of the interior region. Such
an objects are thus direct analogues of the ``black string'' gravitational
configurations reported recently in the literature. 

\vfill

\noindent $^{*})$ On leave from: Department of Theoretical Physics, 
Moscow State University, 117234 Moscow, Russia.

\eject  

\sectio{\bf Introduction}
\medskip

The interest to the classical solutions of the Yang-Mills gauge theory has
been recently revived on the basis of deriving correspondences between the
gauge models with internal symmetry group and the gauge theories of gravity.
There is a wide variety of such maps [1-6] which permit to reformulate the 
Yang-Mills theory in terms of an ``effective'' gravitational Einstein or, 
in general, Einstein-Cartan theory. In particular, the use of such 
reformulations enabled a number of authors [7,8] to find the spherically 
symmetric solutions for the Yang-Mills equations which are analogous to 
the black hole gravitational configurations. Quite recently the study of
cylindrically symmetric gravitational fields has resulted in revealing the
``black string'' solutions [9] which possess a similar property providing
a confinement of a classical particle to the inner region under the 
cylindrical horizon. In the present paper we are using the mapping 
technique to derive the analogous exact cylindrically symmetric solutions 
in the Yang-Mills gauge theory. 

To recall, in the previous studies of the classical Yang-Mills theory the 
attention was paid mainly to the regular solutions with finite energy or/and 
action (a good review of earlier work is given in [10]). However it was 
proven [11] that the finite energy gauge field configurations cannot form
bound states with fermion particles and thus do not possess confining 
properties. It was proposed in [11] that the infinite energy configurations
should be analyzed instead. The new spherically symmetric solutions [7,8]
are all singular and have infinite energy, which indeed yields the confining 
properties analogous to that of a black hole. [As a matter of fact, these 
solutions are not really new, both authors [7,8] have rediscovered the 
results of two papers of Protogenov [12-13]; however their discussion of 
classical confinement is a new development, see also [14]]. 

There was already some interest to more general symmetries, in particular
L. Witten [15] was the first to establish a direct correspondence between 
the axially symmetric Einstein solutions and the static axially symmetric 
self-dual Yang-Mills gauge fields. More recent, partly overlapping, discussion 
is presented in [16]. A cylindrically symmetric case was analyzed in [17], 
which was treated as a leading approximation for the toroidal localized 
configurations.

In the present paper we show that the singular solutions with the confining
properties exist not only for the spherical symmetry [2,8] but 
also for a cylindrically symmetric case. The general interest to this case 
is motivated by the studies of the classical string-like configurations, in
particular in cosmology. When however a periodic conditions are imposed on 
the symmetry axis coordinate, one finds an approximate toroidal solutions. 
Compact toroidal configurations of gluonic and quark matter had attracted 
much attention in the literature (see, e.g., [17,24] and references
therein). It is also worthwhile to mention that the study of a cylindrically
symmetric case is often useful as a preliminary step in treating a more
general axial symmetric problem. We use the mapping from the $SU(2)$ 
Yang-Mills theory into the effective Einstein-Cartan gravity for the 
formulation of the general problem and for the analysis of the properties 
of solutions obtained. An infinite family of solutions is discovered which 
are labeled by two real parameters. 

\medskip
\sectio{\bf Yang-Mills theory mapped into a gauge gravity model}
\medskip

Let us consider the $SU(2)$ gauge theory on the flat Minkowski space-time 
$M$. It will be convenient to use the coordinate-free formulation in terms of
the exterior form language. As in [8] we will study the model of the 
coupled $SU(2)$ gauge field and a triplet of massless Higgs scalar fields 
$\phi^a$, i.e. what is called the Yang-Mills-Higgs theory in the 
Prasad--Sommerfield limit. The Lagrangian is 
$$
L_{\rm YMH}=-{1\over 2}(F^{a}\wedge{\ }^{\ast}F_{a} + D\phi^{a}
\wedge{\ }^{\ast}D\phi_{a}),\eqno(\z)
$$
where $\ast$ denotes the four-dimensional Hodge dual.

In order to describe the mapping from the 
internal gauge theory into an ``effective'' gravity form, we need to make
a formal $(1+3)$ decomposition. Hence we will distinguish the global time 
coordinate $x^0 =t$ from the rest of coordinates $x^{i},\ i=1,2,3,$ which 
parametrize the flat Euclidean three-dimensional space $\underline{M}$ (so 
that $M = R\times\underline{M}$). The (pseudo)--Riemannian structure is 
introduced on $M$ by the Minkowski metric $g$ with the Lorentzian signature 
$(-,+,+,+)$, and in accordance with the product structure we write the line 
element as
$$
ds^2 = - dt\otimes dt + \underline{ds}^2 = 
- dt\otimes dt + \underline{g}_{ij}dx^{i}\otimes dx^{j}\eqno(\z)
$$
with the positive definite 3-metric $\underline{ds}^2 =\underline{g}_{ij}dx^{i}
\otimes dx^{j}$. The latter is flat: a coordinate transformation exist
from $x^i$ to $x'^i$ in which $\underline{g'}_{ij}=\delta_{ij}$. Hereafter 
the underline denotes the three-dimensional quantities and structures on 
$\underline{M}$. 

Developing the $(1+3)$--decomposition (see, e.g. [18]) of the $SU(2)$ 
Yang-Mills theory defined on $M$, one writes the potential 1--form and the
field strength 2-form as 
$$
A^a = dtA_0^a +\underline{A}^a ,\quad\quad
F^a =  dt\wedge E^a + B^a .\eqno(\z)
$$
The `electric' piece of the Yang-Mills field strength reads
$$
E^a:=\underline{{\buildrel .\over A}}{}^a - \underline{D}A_{0}^a =
D_{t}\underline{A}^a - \underline{d}A_{0}^a \eqno(\z)
$$
with $\underline{{\buildrel .\over A}}=\partial_{0}{\underline{A}}$ as the 
time derivative, and the covariant derivatives defined by:
$$
D_t\xi^a :=\dot{\xi}^a +\varepsilon^{a}{}_{bc}A_{0}^b \xi^c,\quad 
\underline{D}\xi^a :=
\underline{d}\xi^a +\varepsilon^{a}{}_{bc}\underline{A}^b \xi^c .\eqno(\z)
$$
The `magnetic' piece of the field strength is 
$$
B^a :=\underline{F}^a =\underline{d}\;\underline{A}^a +
{1\over 2}\varepsilon^{a}{}_{bc}\underline{A}^b\wedge\underline{A}^c.\eqno(\z)
$$

The Lagrangian (2.1) is decomposed as
$$
L_{\rm YMH}={1\over 2}dt\wedge\left(- B^{a}\wedge{\ }^{\underline{\ast}}B_{a}+ 
E^{a}\wedge{\ }^{\underline{\ast}}E_{a} +
D_{t}\phi^{a}\wedge{\ }^{\underline{\ast}}
D_{t}\phi_{a} - \underline{D}\phi^{a}\wedge{\ }^{\underline{\ast}}
\underline{D}\phi_{a}\right),\eqno(\z)
$$
where $\underline{\ast}$ stands for {\it the 3-dimensional Hodge dual},
defined by the flat 3--metric $\underline{g}$. 

We now transform the Yang--Mills theory into the form of effective gauge 
gravity model. There are several ways to do this [1-5], 
but here we consider the mapping due to Lunev [2,7] which is 
defined when a fixed background (e.g., the Minkowskian) geometry is present.
As a first step, we rewrite the Yang-Mills theory in the first order form. 
This is straightforward after the introduction of two auxiliary fields, 
1--form $\Theta_{a}$ and 2--form $\pi_{a}$ (both transform covariantly under 
the action of the gauge group):
$$
L_{\rm YMH}=dt\wedge\Big( - \Theta_{a}\wedge B^{a} + {1\over 2}\Theta_{a}
\wedge{\ }^{\underline{\ast}}\Theta^{a} 
$$
$$ 
+ \pi_{a}\wedge E^{a} - {1\over 2}\pi_{a}\wedge{\ }^{\underline{\ast}}\pi^{a}
+ D_{t}\phi^{a}\wedge
{\ }^{\underline{\ast}}D_{t}\phi_{a} - \underline{D}\phi^{a}\wedge
{\ }^{\underline{\ast}}\underline{D}\phi_{a}\Big).\eqno(\z)
$$
The basic mapping is formulated as follows: for every three--dimensional 
Yang--Mills configuration $(\underline{A}^a , \Theta^a )$ we define 
{\it the three--dimensional Riemann--Cartan effective} geometry by
$$
\underline{\Gamma}^{ab} = A^{c}\varepsilon_{c}{}^{ab},\quad\quad
\underline{\vartheta}^{a}=\Theta^{a}.\eqno(\z)
$$
Then the magnetic strength (2.6) is mapped into the three-dimensional
curvature two-form $\underline{R}^{ab}$ and first term in the Lagrangian 
(2.8) is just the standard Hilbert--Einstein gravitational term,
$$
-\Theta_{a}\wedge B^{a}=-{1\over 2}\eta_{ab}\wedge\underline{R}^{ab},
$$
while the second term represents a sort of a generalized ``cosmological 
constant''. Except for ($\underline{\vartheta}^a, \underline{\Gamma}^{ab}$) 
all the rest variables (the Yang-Mills field time component $A^{a}_{0}$, the 
variable $\pi^a$, and $\phi^a$) then should be treated as the effective matter 
described by the second line in (2.8). Let us derive the effective 
``gravitational'' field equations. Independent variation of (2.8) 
with respect to $\underline{\vartheta}^a$ and $\underline{\Gamma}^{ab}$ yield, 
respectively,
$$
\underline{R}^{ab} = \varepsilon^{abc}\ {}^{\underline{\ast}}
\underline{\vartheta}_{c}, \eqno(\z)
$$
$$
\underline{T}^{a} = -D_{t}\pi^{a}- \varepsilon^{a}{}_{bc}
\phi^{b\underline{\ast}}\underline{D}\phi^c,\eqno(\z)
$$
where $\underline{T}^{a}$ is the effective torsion two-form. Notice the 
absence of a ``matter source'' in the effective Einstein equation (2.10). 
As one notices (correcting the statements made in [2,7]), in 
general case the effective spatial geometry is the
Riemann-Cartan one: torsion (2.11) is non trivial for non-static
configurations, and both $A_{0}$ and Higgs scalars contribute to it.
Variation of (2.8) with respect to the ``matter'' fields yields, 
respectively,
$$
\underline{D}\pi^a =-\varepsilon^{a}{}_{bc}\phi^{b\underline{\ast}}D_{t}\phi^c,
\quad\quad  E^a = ^{\underline{\ast}}\pi^a ,\eqno(\z)
$$
while the scalar multiplet $\phi^a$ satisfies the generalized Klein-Gordon 
field equation,
$$
D_{t}{\ }^{\underline{\ast}}D_{t}\phi^a - \underline{D}{\ }^{\underline{\ast}}
\underline{D}\phi^{a}=0.\eqno(\z)
$$

\medskip
\sectio{\bf Cylindrically symmetric configurations}
\medskip

We are going to discuss the exact solutions with a cylindrical symmetry. 
Hence we cover the 3-space $\underline{M}$ by the cylindrical coordinate 
system, $x^{i}=\{\rho, \theta, z\}$, with the background 3-metric in its 
standard form
$$
\underline{ds}^2 = d\rho^2 + \rho^2 d\theta^2 + dz^2.\eqno(\z)
$$

We look for the static cylindrically symmetric solutions of the Yang-Mills 
equations. In the effective gravity theory (2.11)-(2.10) we should 
search for the effective frame and connection. The most general cylindrically 
symmetric ansatz reads
$$
\underline{\vartheta}^a = 
\pmatrix{A(\rho)d\rho\cr B(\rho)d\theta\cr C(\rho)dz\cr},\quad\quad
\underline{\Gamma}_{a}{}^{b} = \pmatrix{0\ & Ud\theta\ &V\rho^{-1}dz\ \cr 
-Ud\theta\ &0\ &0\ \cr -V\rho^{-1}dz\ &0\ &0\ \cr}\eqno(\z)
$$
Hereafter we use the self-evident matrix notation. The functions 
$U=U(\rho), V=V(\rho), A, B, C$ determine the gauge field configuration. 
Extra $\rho$ factors are introduced in (3.2) for later convenience.
It seems worthwhile to mention that in the effective gravity framework it is
quite straightforward to derive the symmetric ansatz for any field variable.
In particular, (3.2) is suggested naturally by the Cartan structure 
equations. The torsion and curvature 2-forms for the Riemann-Cartan gauge
fields (3.2) are as follows,
$$
T^a = \pmatrix{0\cr (B' + AU)d\rho\wedge d\theta\cr 
(C' + AV\rho^{-1})d\rho\wedge dz\cr},\eqno(\z)
$$
$$
R_{a}{}^{b} = 
\pmatrix{0\ &U'd\rho\wedge d\theta\ &(V\rho^{-1})'d\rho\wedge 
dz\ \cr -U'd\rho\wedge d\theta\ &0\ &-UV\rho^{-1}d\theta\wedge dz\ \cr 
-(V\rho^{-1})'d\rho\wedge dz\ &UV\rho^{-1}d\theta\wedge dz\ &0\ \cr}\eqno(\z)
$$

For static case the equations (2.11), (2.12) and (2.13) can be 
rewritten, excluding $\pi^a$, as
$$
T^{a} = \varepsilon^{a}{}_{bc}A^{b}_{0}{\ }^{\underline{\ast}}
(\underline{D}A^{c}_{0}) -\varepsilon^{a}{}_{bc}\phi^{b}{\ }^{\underline{\ast}}
(\underline{D}\phi^{c}),\eqno(\z)
$$
$$
\underline{D}{\ }^{\underline{\ast}}\underline{D}A^{a}_{0}=0,\eqno(\z)
$$
$$
\underline{D}{\ }^{\underline{\ast}}\underline{D}\phi^{a}=0,\eqno(\z)
$$
respectively. The cylindrically symmetric ansatz for $A^{a}_{0}$ and $\phi^a$ 
reads
$$
A^{a}_{0} = {1\over \rho}\pmatrix{W(\rho)\cr 0\cr 0\cr},\quad\quad
\phi^{a} = {1\over \rho}\pmatrix{H(\rho)\cr 0\cr 0\cr}.\eqno(\z)
$$

The Hodge duality operation defined by the flat metric (3.1) can be
summarized in a matrix notation as
$$
\underline{\ast}\pmatrix{d\rho\cr d\theta\cr dz\cr} = \pmatrix{\rho
d\theta\wedge dz\cr \rho^{-1}dz\wedge d\rho\cr \rho d\rho\wedge d\theta\cr}.
\eqno(\z)
$$

Substituting (3.8), (3.2) and (3.9) into (3.6) and (3.7), we find
$$
\rho^2 W'' - \rho W' = W\left(U^2 + V^2 - 1\right),\eqno(\z)
$$
$$
\rho^2 H'' - \rho H' = H\left(U^2 + V^2 - 1\right).\eqno(\z)
$$

Using (3.3), (3.8) and (3.9), one transforms the torsion
(``Cartan'') equation (3.5) and the Einstein equation (2.10) to 
$$
\rho^2 U'' - \rho U' = U\left(V^2 + H^2 - W^2\right),\eqno(\z)
$$
$$
\rho^2 V'' - \rho V' = V\left(U^2 + H^2 - W^2 - 1\right).\eqno(\z)
$$
These equations (3.10)-(3.13) form a closed system of second order 
non-linear equations for the functions $U, V, W, H$. 
The effective metric coefficients $A, B, C$ are constructed from them via
$A=UV\rho^{-2}, B=-\rho(V/\rho)', C=-U'/\rho$.

\medskip
\sectio{\bf Exact solutions with confining properties}
\medskip

Noticing that the general structure of the system (3.10)-(3.13) 
is close to that of the spherically symmetric problem
in the Prasad-Sommerfield limit [19], we look for the general 
non-degenerate solution for $V, H, W$ in the form
$$
V=K(\rho),\quad H=K(\rho)\cosh\gamma, \quad W=K(\rho)\sinh\gamma,\eqno(\z)
$$
where $\gamma$ is an arbitrary constant. Using (4.1), one reduces
(3.10)-(3.13) to the following system of coupled 
equations for two unknown functions $K, U$:
$$
\rho^2 K'' - \rho K' = K\left(K^2 + U^2 - 1\right),\eqno(\z)
$$
$$
\rho^2 U'' - \rho U' = 2UK^2.\eqno(\z)
$$
One immediately notices some resemblance of these equations to the spherically
symmetric equations in the Prasad-Sommerfeld limit. However the second terms
on the l.h.s. produce essential difference. In particular, the analysis of
the power series expansions at zero and at infinity shows that the system
(4.2)-(4.3) does not admit analytical solutions in which the function $K$
goes to $\pm 1$ at zero or at infinity. [To check our conclusions, we have 
used the REDUCE-based [20] computer algebra system GRG [21] in calculations].

Integration of (4.2)-(4.3) is simplified greatly when one notices that
this system is a consequence of the {\it first order} system,
$$
\rho K' - K = \varepsilon K U,\eqno(\z)
$$
$$
\rho U' = \varepsilon K^2, \eqno(\z)
$$
where $\varepsilon^2 =1$. As one can see, (4.4)-(4.5) possesses a
first integral 
$$
(\varepsilon U + 1)^2 - K^2 = C, 
$$
with the help of which the final integration of (4.5) or (4.4) is 
straightforward. We have three cases, depending on the value of the constant
$C$.

For $C=0$ we find
$$
K= {\varepsilon_1\over {\log\left({\rho\over \rho_{0}}\right)}},\quad
U= \varepsilon_2\left( 1 + {1\over {\log\left({\rho\over 
\rho_{0}}\right)}}\right),\eqno(\z)
$$
where $\varepsilon_{1,2}^2 =1$, and $\rho_{0}$ is an integration constant.

For negative constant $C=-n^2$ we get 
$$
K_{(n)}= {\varepsilon_1 n\over\cos\left(n\ {\log\left({\rho\over \rho_{0}}
\right)}\right)},\quad
U_{(n)}= \varepsilon_2 \left( -1 + n\tan\left(n\ {\log\left({\rho\over 
\rho_{0}}\right)}\right)\right),\eqno(\z)
$$
while for positive constant $C=n^2$ one finds
$$
K_{(n)}=
{2\varepsilon_{1}n\left({\rho\over\rho_{0}}\right)^n\over 
{\left({\rho\over\rho_{0}}\right)^{2n} - 1}},\quad 
U_{(n)}=\varepsilon_2\left(1+n+{2n\over{\left
({\rho\over\rho_{0}}\right)^{2n} - 1}}\right).\eqno(\z)
$$
Below for definiteness we will consider the plus signs $\varepsilon_1=
\varepsilon_2 = +1$ in these solutions.

The classical solutions (4.7) and (4.8) are parametrized by the two
{\it real} constants: $n$ and $\rho_0$. In fact, also any complex $n$ is 
formally admissible and a complex integration constant $\rho_0$, but then 
the $SU(2)$ gauge fields are also complex and their physical interpretation 
is unclear. Solutions (4.8) with non-integer $n$ is not analytical at zero 
and infinity. Negative $n$ does not give anything new, it is easy to see that
$$
K_{(-n)}=K_{(n)},\quad\quad U_{(-n)}=U_{(n)}.
$$
Trivial $n=0$ gives also a solution, which however does not reduce to (4.6).

All the solutions are singular. The fields (4.7) have infinite number of
singular points, while the configurations (4.6) have logarithmic, and
(4.8) a simple pole behavior at a point $\rho = \rho_0$,
$$
K_{(n)}\vert_{\rho\rightarrow\rho_0}= U_{(n)}\vert_{\rho\rightarrow\rho_0}= 
{\rho_0\over {\rho - \rho_0}}.\eqno(\z)
$$
This is most easily seen for the integer $n$, then one can use the explicit
formula $(r^{2n} - 1)=(r-1)(r^{2n-1}+r^{2n-2}+...+r+1)$ (with $r=\rho/\rho_0$).
But (4.9) is valid also for arbitrary non-integer $n$. This is the
same singularity which is typical for the spherically symmetric solutions
with confining properties [7,8]. However unlike the spherical
case the Yang-Mills potentials and scalar fields (3.2), (3.8)
are regular at the origin $\rho=0$. The rest of the paper is devoted to
the discussion of the solutions (4.8).

Without the loss of generality we will put $\rho_{0}=1$ in our subsequent 
study of the solution properties. Let us analyze the particular solution 
with $n=1$ in greater detail. 

The effective ``gravitational'' fields, which completely characterize the
Yang-Mills field configuration, are calculated straightforwardly. When $n=1$
we find for the torsion and curvature the following expressions, respectively,
$$
T^a = {-8\rho\over (\rho^2 -1)^3}\pmatrix{0\cr d\rho\wedge d\theta\cr 
d\rho\wedge dz\cr},\eqno(\z)
$$
$$
R_{a}{}^{b} = {-4\rho\over (\rho^2 -1)^2}
\pmatrix{0\ &d\rho\wedge d\theta\ &d\rho\wedge dz\ \cr -d\rho\wedge d\theta\ 
&0\ &\rho d\theta\wedge dz\ \cr 
-d\rho\wedge dz\ &-\rho d\theta\wedge dz\ &0\ \cr}\eqno(\z)
$$
Recall that the Riemann-Cartan curvature (4.11) is in fact the magnetic
part of the Yang-Mills field strength. 

The effective three-metric components $A, B, C$ are also easily calculated to
give for the line element
$$
\underline{ds}^{2}_{\rm eff} = {16\over (\rho^2 -1)^4}(\rho^2 d\rho^2 + 
\rho^4 d\theta^2 + dz^2).\eqno(\z)
$$

As we see, all the fields are zero at the symmetry axis, become infinite
at a distance $\rho=1$ and then fastly fall off to zero at large distances
from the axis. Solutions with $n>1$ have the same general properties, only
they decrease more quickly when $\rho\longrightarrow 0$ and when 
$\rho\longrightarrow\infty$. The physical interpretation of such a Yang-Mills
configuration is clear: A region of space inside the tube of the radius
$\rho=1$ is separated by an infinite potential barrier from the outside
space, providing a classical confinement of any matter with gauge charges
in string-like structure. The absence of the gauge field at $\rho =0$ 
is a classical counterpart of asymptotic freedom, since no force is acting
on the gauge charges on the axis. One thus may call the family (4.8)
a confining string solutions. 

Due to the simple pole singularity (4.9) the total energy of all the
solutions diverges, and it seems necessary to introduce a proper cutoff like
with the formally infinite energy Coulomb solution, following the suggestion
of [8]. It should be stressed however that the confining property of 
the classical solutions (4.8) is a direct consequence of their 
singularity. 

\medskip
\sectio{\bf Quantum particle confinement}
\medskip

Analogously to the spherically symmetric case [7,14] one can 
demonstrate that quantum particles with gauge charge are indeed confined
inside the cylindrical domain near the symmetry axis. Let us consider the
isospin 1/2 scalar particle in an external gauge field (4.8). For
simplicity we will choose $\gamma=0$ in (4.1) (thus eliminating the
time component of the gauge field), and $n=1$ and set the integration 
constant $\rho_{0}=1$ (for other values of $n$ and $\rho_{0}$ the results 
are qualitatively the same).

The dynamics of a 2-component scalar field $\Psi^{A}, A=1,2$ with mass $M$ 
in the external gauge field is described by the Klein-Gordon equation
$$
{\buildrel .. \over \Psi}{}^A  - {\ }^{\underline{\ast}}\underline{D}
{\ }^{\underline{\ast}}\underline{D}\Psi^{A} + M^2\Psi^{A}=0,\eqno(\z)
$$
where the covariant derivative is 
$$
\underline{D}^{A}_{B}:= \underline{d}\ \delta^{A}_{B} - 
{i\over 2}\underline{A}^{a}\left(\sigma_{a}\right)^{A}_{B},
$$
and $\sigma_{a}$ are $2\times 2$ Pauli matrices.

Substituting $n=1$ (4.8) gauge field configuration into (5.1),
and looking for the finite energy solutions
$$
\Psi^{A}=e^{-itE}\psi^{A}(\rho,\theta,z),
$$
we find
$$
-\left({1\over \rho}{\partial\over\partial\rho}\left(\rho{\partial\psi^{A}\over
\partial\rho}\right) + {1\over \rho^2}{\partial^2\psi^{A}\over\partial\theta^2}
 + {\partial^2\psi^{A}\over\partial z^2}\right)
$$
$$ 
+ \left(M^2 - E^2 + {{\rho^2 + 1}\over(\rho^2 -1)^2} - 
{2\over(\rho^2 -1)}\Lambda\right)\psi^{A} =0,\eqno(\z)
$$
where we denote a linear differential operator
$$
\Lambda:= -i\left(\sigma^{2}{\partial\over\partial z} - 
\sigma^{3}{\partial\over\partial \theta}\right).\eqno(\z)
$$
It is easy to see that the operators
$$
-i{\partial\over\partial \theta},\quad -i{\partial\over\partial z},\quad 
\Lambda
$$
commute with each other and with the differential operator in (5.2).
Hence one can look for the solution which is a common eigenstate of all the
operators, and separate variables as 
$$
\psi^{A}={1\over\sqrt{\rho}}\varphi_{ml\lambda}(\rho)e^{imz +il\theta}
\psi^{A}_{\lambda},\eqno(\z)
$$
where constants $m$ and $l$ are the eigenvalues of $-i\partial_{z}$ and
$-i\partial_{\theta}$, respectively, and $\psi^{A}_{\lambda}$ is the 
eigenfunction 
$$
\Lambda\psi^{A}_{\lambda}=\lambda\psi^{A}_{\lambda}
$$
with the eigenvalues $\lambda=\pm\sqrt{m^2 + l^2}$. 

Substituting (5.4) into (5.2) we find for the radial function the
one-dimensional stationary Schr\"odinger equation
$$
\left(-{d^2\over d\rho^2} + V(\rho)\right)\varphi_{ml\lambda}=
\mu^2\varphi_{ml\lambda}.\eqno(\z)
$$
with the potential
$$
V(\rho)={{\rho^2 +1}\over(\rho^2 -1)^2} - 
{2\lambda\over(\rho^2 -1)} - {(l^2 + {1\over 4})\over\rho^2} ,\eqno(\z)
$$
and eigenvalue
$$
\mu^2 = E^2 - M^2 + m^2.
$$
The infinite potential barrier at $\rho=1$ with the leading term 
$V\sim {2\over (\rho -1)^2}$ is known to be completely impenetrable [22]
(the exact form of the eigenfunctions in (5.5) is not important). Thus
all quantum particles with a gauge charge cannot move out of the ``thick
string'' region $0\leq\rho\leq 1$, providing a picture of confinement.
The same conclusion is valid also for the Dirac spinor particles.

\medskip
\sectio{\bf Discussion and conclusion}
\medskip

In this paper we obtained a family of new exact cylindrically symmetric 
solutions for the $SU(2)$ gauge Yang-Mills theory. Like the earlier reported
spherically symmetric solutions, these are also singular and can thus provide 
a mechanism for the classical confinement. This is another demonstration of 
the fruitfulness of deriving analogies and constructing direct mappings 
between the gauge theories of internal symmetry groups and gravity theory.

It seems necessary to mention certain problems and prospects for the results
obtained. First of all, QCD works not with $SU(2)$ group but with $SU(3)$. 
Straightforward analysis shows that there exist a generalization of the 
above results on the  $SU(3)$ case, which arise from the embedding of $SU(2)$ 
into $SU(3)$. However the geometrical meaning of the mapping between the 
gauge model and gravity is at the moment unclear to us, although one can
point on the recent works [3,23] where the attempts were made to prove the 
existence of the reasonable generalization of the mapping between $SU(N), 
N\geq 3,$ gauge theory and gravitation, thus demonstrating that the 
coincidence of the $SU(2)$ group dimension with the number of spatial 
coordinates is accidental and plays no decisive role. The arising 
effective spatial geometry is necessarily non-Riemannian in this case. 
Another point which deserves special attention is the physical interpretation
of the infinite energy of the confining solutions. As we already mentioned,
finite-energy configurations cannot provide the binding of gauge charges.
However a regularization similar to the one which is applied to the
(originally infinite-energy) meron solutions might be necessary in our
case. The work is in progress in this direction.

A possible physical importance of the new solutions would be a development
of the toroidal bag constituent model for the hadrons. Recently an analogous
spherical bag model based on the classical singular solutions was discussed in 
[14]. The toroidal glueballs were analyzed previously, e.g., in 
[24], and later in [17] an attempt was made to reformulate this 
model within the cylindrically symmetric approximation to the Yang-Mills 
toroidal configurations. Such an approximation arises naturally from an exact
cylindrical solutions by imposing a periodicity condition on the $z$ coordinate.
Some remarks are in order about the papers [17]. Using a similar but
somewhat different from (3.2) ansatz for the cylindrically symmetric
Yang-Mills field, the authors of [17] failed to describe an exact classical 
solution. They however noticed the possibility of a simple pole singularity 
of the type (4.9) and tried to avoid it by assuming that the integration 
constant analogous to $\rho_{0}$ is complex, at the same time speculating
that a smooth matching of such a solution with the real analytical power 
series solutions at zero and infinity exists. As it is clearly shown in our 
paper, no such matching exists and the only possibility to avoid the 
singularity for real $\rho$ is to make a solution complex on all the $\rho$ 
axis. Such a complex solution, although formally admissible, will be most 
probably unphysical. Our results thus provide corrections and generalization 
of the work [17].

It may turn out certainly that the true explanation of the QCD confinement 
involves purely quantum arguments which are unrelated to the speculative, to 
an extent, calculations based on the new spherically symmetric [7,8]
and cylindrically symmetric (4.8) Yang-Mills solutions. Nevertheless,
it seems worthwhile to notice once again the power and flexibility of the
classical Yang-Mills theory which provides an alternative geometrical 
mechanisms for understanding, maybe at least partly, of such intriguing 
problems as the confinement. 

\bigskip
{\bf Acknowledgments}. 

I would like to thank Friedrich W. Hehl, Frank Gronwald and Franz Schunck 
for helpful remarks and discussions of the results obtained. This work was 
supported by the Deutsche Forschungsgemeinschaft (Bonn) grant He 528/17-1. 

\medskip
\eject
{\bf References}
\medskip
\newref[1]
K. Johnson: {\it The Yang--Mills ground state}, in: {\sl QCD -- 20
Years Later, Aachen 1992}, P.M. Zerwas and H.A. Kastrup, eds., Vol.2
(World Scientific, Singapore 1993) pp.795; 
D.Z. Freedman, P.E. Haagensen, K. Johnson, and J.I. Latorre, {\it The
hidden spatial geometry of non-Abelian gauge theories}, {\sl Preprint
CERN-TH.7010/93};
D.Z. Freedman, and R.R. Khuri, {\sl Phys. Lett.} {\bf B329} (1994) 263.
\newref[2]
F.A. Lunev, {\sl Phys. Lett.} {\bf B295} (1992) 99;
F.A. Lunev, {\sl Theor. and Math. Phys.} {\bf 94} (1993) 48;
F.A. Lunev, {\sl Phys. Lett.} {\bf B314} (1993) 21;
F.A. Lunev, {\sl Mod. Phys. Lett.} {\bf A9} (1994) 2281.
\newref[3]
M. Bauer, D.Z. Freedman, and P.E. Haagensen, {\sl Nucl.Phys.} {\bf B428}
(1994) 147;
M. Bauer and D.Z. Freedman, {\sl Nucl.Phys.} {\bf B450} (1995) 209.
\newref[4]
E.M. Mielke, Yu.N. Obukhov, and F.W. Hehl, {\sl Phys. Lett.} {\bf A192} (1994)
153.
\newref[5]
P.E. Haagensen and K. Johnson, {\sl Nucl. Phys.} {\bf B439} (1995) 597.
\newref[6]
V. Radovanovic and Dj. \v{S}ija\v{c}ki, {\sl Class. Quantum Grav.} {\bf 12} 
(1995) 1791.
\newref[7]
F.A. Lunev, {\sl Phys. Lett.} {\bf B311} (1993) 273.
\newref[8]
D. Singleton, {\sl Phys. Rev.} {\bf D51} (1995) 5911.
\newref[9]
J.P.S. Lemos, {\it Cylindrical black hole in general relativity}, {\sl 
Preprint, Lanl e-archive gr-qc/9404041} (1994) 13 p.; J.P.S. Lemos and
V.T. Zanchin, {\it Rotating Charged Black String and Three Dimensional 
Black Holes}, {\sl Preprint, Lanl e-archive hep-th/9511188} (1995) 68 p.
\newref[10]
A. Actor, {\sl Rev. Mod. Phys.} {\bf 51} (1979) 461.
\newref[11]
J.H. Swank, L.J. Swank, and T. Dereli, {\sl Phys. Rev.} {\bf D12} (1975) 1096.
\newref[12]
A.P. Protogenov, {\sl Phys. Lett.} {\bf B67} (1977) 62.
\newref[13]
A.P. Protogenov, {\sl Phys. Lett.} {\bf B87} (1979) 80. 
\newref[14]
D. Singleton and A.Yoshida, {\it A general relativistic model for confinement
in SU(2) Yang-Mills theory}, {\sl Preprint, Lanl e-archive hep-th/9505160} 
(1995) 17 p.
\newref[15]
L. Witten, {\sl Phys. Rev.} {\bf D19} (1979) 718.
\newref[16]
D. Singleton, {\it Axially symmetric solutions for SU(2) Yang-Mills theory},
{\sl Preprint, Lanl e-archive hep-th/9502116} (1995) 15 p.
\newref[17]
S.M. Mahajan and P.M. Valanju, {\sl Phys. Rev.} {\bf D35} (1987) 2543;
S.M. Mahajan and P.M. Valanju, {\sl Phys. Rev.} {\bf D36} (1987) 1500.
\newref[18]
E.W. Mielke, {\sl Ann. Phys. (USA)} {\bf 219} (1992) 78. 
\newref[19]
M.K. Prasad and C.M. Sommerfield, {\sl Phys. Rev. Lett.} {\bf 35} (1975) 760.
\newref[20]
D. Stauffer, F.W. Hehl, N. Ito, V. Vinkelmann, and J.G. Zabolitzky, {\it 
Computer simulation and computer algebra}, 3rd ed. (Springer, Berlin, 1993).
\newref[21]
V.V. Zhytnikov, I.G. Obukhova, and S.I. Tertychniy, {\it Computer Algebra 
Program for Gravitation}, in: {\sl Abstracts of contributed papers of the 
GR-13} (Huerta Grande, Cordoba, Argentina, 1992) p. 309;
V.V. Zhytnikov, {\it GRG. Computer Algebra  System for Differential
Geometry, Gravity and Field Theory. Version 3.1. User's manual} (Moscow, 1991) 
108 pp.
\newref[22]
J. Dittrich and P. Exner, {\sl J. Math. Phys.} {\bf 26} (1985) 2000.
\newref[23]
F.A. Lunev,{\it Reformulation of QCD in the language of general relativity},
{\sl Preprint, Lanl e-archive hep-th/9503133} (1995) 25 p.
\newref[24]
D. Robson, {\sl Z. Physik} {\bf C3} (1980) 199.
\bye